\documentclass[twocolumn]{aastex6}

\usepackage{geometry}                % See geometry.pdf to learn the layout options. There are lots.
\usepackage{graphicx}
\usepackage{amsmath,amssymb}
\usepackage{epstopdf}

\usepackage {pstricks}

\begin{document}
%\maketitle

\title{Pre-nebular Light curves of Type I supernovae}

\author{W. David Arnett\altaffilmark{1} 
%\email{ wdarnett@gmail.com}
}

\author{Christopher Fryer\altaffilmark{2}
}

\author{Thomas Matheson\altaffilmark{3}
}

\altaffiltext{1}{Steward Observatory, University of Arizona, 
933 N. Cherry Avenue, Tucson AZ 85721}

\altaffiltext{2}{Los Alamos National Laboratory, Los Alamos NM}

\altaffiltext{3}{National Optical Astronomy Observatory, Tucson AZ}

% Activate to display a given date or no date
{\bf DRAFT FROM \today}

\begin{abstract}
We compare analytic predictions of supernova light curves with recent high quality data from  SN2011fe (Ia), from KSN2011b (Ia), 
and the Palomar Transient Factory (PTF) and the La Silla-QUEST variability survey (LSQ) (Ia). 
 Because of the steady, fast cadence of observations, KSN2011b provides unique new information on SNe Ia: the smoothness of the light curve, which is consistent with significant large-scale mixing during the explosion, possibly due to 3D effects (e.g., Rayleigh-Taylor instabilities), and provides support for a slowly-varying leakage (mean opacity). 
For a more complex light curve (SN2008D, SNIb), we separate the luminosity due to multiple causes and indicate the possibility of a radioactive plume.
The early rise in luminosity is shown to be affected by the opacity (leakage rate) for thermal and non-thermal radiation.
A general derivation of Arnett's rule again shows that it depends upon {\em all} processes heating the plasma, not just radioactive ones, so that SNe Ia will differ from SNe Ibc if the latter have multiple heating processes.
\end{abstract}

\section{Introduction}

Supernovae whose luminosity is primarily powered by the decay chains 
$\rm ^{56}Ni(e^-,\nu){^{56}Co}$,
$\rm ^{56}Co(e^-,\nu){^{56}Fe}$ and $\rm ^{56}Co(,e^+\nu){^{56}Fe}$ have generic features. These are prominent in  SNe Ia, and appear in SNe Ibc, that is, in both thermonuclear supernovae and those core-collapse supernovae which have lost their hydrogen envelopes  (e.g., \cite{a82a,wda96}). 

We compare theoretical light curves \citep{a82a,pe00a,pe00b} to the
best-observed typical  SN Ia to date, SN2011fe in M101, the
Pinwheel galaxy \citep{sn2011fe}, as well as to KSN2011b, the best of
three SNe~Ia (KSN2011b,c; KSN2012a) detected at early times by the Kepler satellite \citep{keplersn}. For these supernovae we have bolometric (or near bolometric) light curves, so we may minimize issues of frequency-dependent atmospheric physics. This approach, which assumes three--dimensional (3D) incomplete mixing during the explosion \citep{am16key}, is thus different from and a natural complement to a 1D, time--dependent stellar atmosphere approach such as \cite{sblondin,dessart,dessart16}. 
%In addition the analytic solutions have very high cadence, which allows a quantitative and precise study of the light curve shapes.

In \S\ref{uniqueness} we introduce the combined problem of parameters and uniqueness, which may be more clearly discused in our analytic framework.
    In \S\ref{feLightcurves} we discuss SN2011fe, including the early light curve.
    In \S\ref{leakage} we discuss leakage of thermal and non-thermal radiation from supernovae in terms of an effective opacity.
    In \S\ref{section-ksn} we discuss KSN2011b, the problem of converting it to a bolometric scale, and show that it has an exceptionally smooth light curve, constraining theoretical models of mixing.
    In \S\ref{rule} we present a more general derivation of Arnett's rule \citep{a79,a82a}.
In \S\ref{sn2008D} we discuss the core-collapse SN2008D (type Ib), and separate the breakout, shock, and radioactive heating parts of the light curve. 
\S\ref{Conclusion} contains our summary.

\section{Uniqueness}\label{uniqueness}
Light curves of Type I supernovae have a characteristic shape  which is due to radioactive heating by gamma-rays and positrons, cooling by expansion, and cooling by radiative loss \citep{a82a,pe00a}. 
The {\em shape} of these light curves is determined by a parameter 
\begin{equation}
y = (2\tau_d \tau_h)^{1 \over 2} /2\tau_{Ni} \sim (\kappa_t M_{ej} /v_{sc})^{1 \over 2},\label{eq.def-y}
\end{equation} 
which is a combination of diffusion time and expansion time, relative to the a reference time (the decay time for $\rm^{56}Ni$), or equivalently the effective opacity times mass ejected divided by the velocity scale. The {\em amplitude} (peak light) is determined by 
\begin{eqnarray}
L&=& \epsilon_{Ni} M_{Ni} M_\odot \Lambda(x,y) \nonumber\\
&= & 2.055 \times 10^{10} L_\odot M_{Ni} \Lambda(x,y) \label{eq.Lxy}
\end{eqnarray} 
where the dimensionless function $\Lambda$ is determined by an integration in time, and $M_{Ni}$ is the mass of $^{56}$Ni in solar units. An estimate of radioactive heating by a mass $M_{Ni}$ at peak light requires an estimate of the time between explosion (synthesis) and peak light. The theory is {\em not} a one-zone model as in \cite{a79}, but involves integration over both space and time by separation of these variables (see \cite{a82a}, Eq.~11 and 48).

The light curves are determined  by five astrophysical parameters, acting in combination 
(Eq.~\ref{eq.def-y} and Eq.~\ref{eq.Lxy})\footnote{The Phillips relation \citep{phillips93} was discovered as a feature of B-band photometry, but may be realized in the bolometric light curves if $y$ increases with $M_{Ni}$, for example. From Fig.~\ref{fig_sn2011fe} we find a ``bolometric $\Delta m_{15}$'' of $\sim 1.0$ mag for SN2011fe.}. 
They are the mass ejected $M_{ej}$, the mass of $\rm^{56}Ni$ ejected\footnote{\cite{tac67} did the first essentially complete reaction network studies of nucleosynthesis in supernova shocks, showing that the radioactive nucleus $\rm ^{56}Ni$ was the dominant product, with smaller amounts of $\rm^{55}Co$ and $\rm^{57}Ni$. The latter have little effect on the early light curve discussed here ($0 \leq t \leq 40$ days after explosion), however they are important diagnostics \citep{ivo2009,roepke}.}  $M_{Ni}$, the initial radius $R(0)$, the effective opacity $\kappa_t$ (a measure of leakage rate for radiant energy, see \S\ref{leakage}), and the explosion energy $E¨_{sn}$.
The initial radius\footnote{ For a typical expansion velocity of $\approx 10^9$ cm s$^{-1}$, little luminosity is produced before times  $\approx 10^6$ s for such small radii.} of the white dwarf is $R(0)\ll 10^{13}$ cm so that its precise value is unimportant for the light curve.
If the remaining four parameters are constrained to be consistent with a thermonuclear explosion of a white dwarf, realistic light curves for SNe Ia result. 

The situation is more complex for SNe Ibc: (1) there is a collapsed remnant which may be active, and (2)  there may be fluid dynamic heating due to interaction of the ejecta  with a mantle or any surrounding gas, for example\footnote{SNe Ia may also interact with the circumstellar medium (CSM); \cite{hamuy}.}.  

This additional physics may imply new parameters, further adding to the problem of uniqueness. 
If we infer $y$ from the shape of the light curve, this only fixes the combination $\kappa_t M_{ej} /v_{sc}$ (not the individual values, for which additional constraints are needed).  It might be a good project to examine spectral estimates of velocity to constrain velocity structure, for example. So far this seems to have been done using $v(t)$, which maps into radius $v(r)$ easily enough, but not mass coordinate $v(m)$ with any precision.  That is required because $E_{KE}= 0.3 v_{sc}^2/M_{ej}$.

\section{Light curves for SN2011fe}\label{feLightcurves}

\begin{figure}[h]
\figurenum{1}
\includegraphics*[angle=-90,scale=0.27]{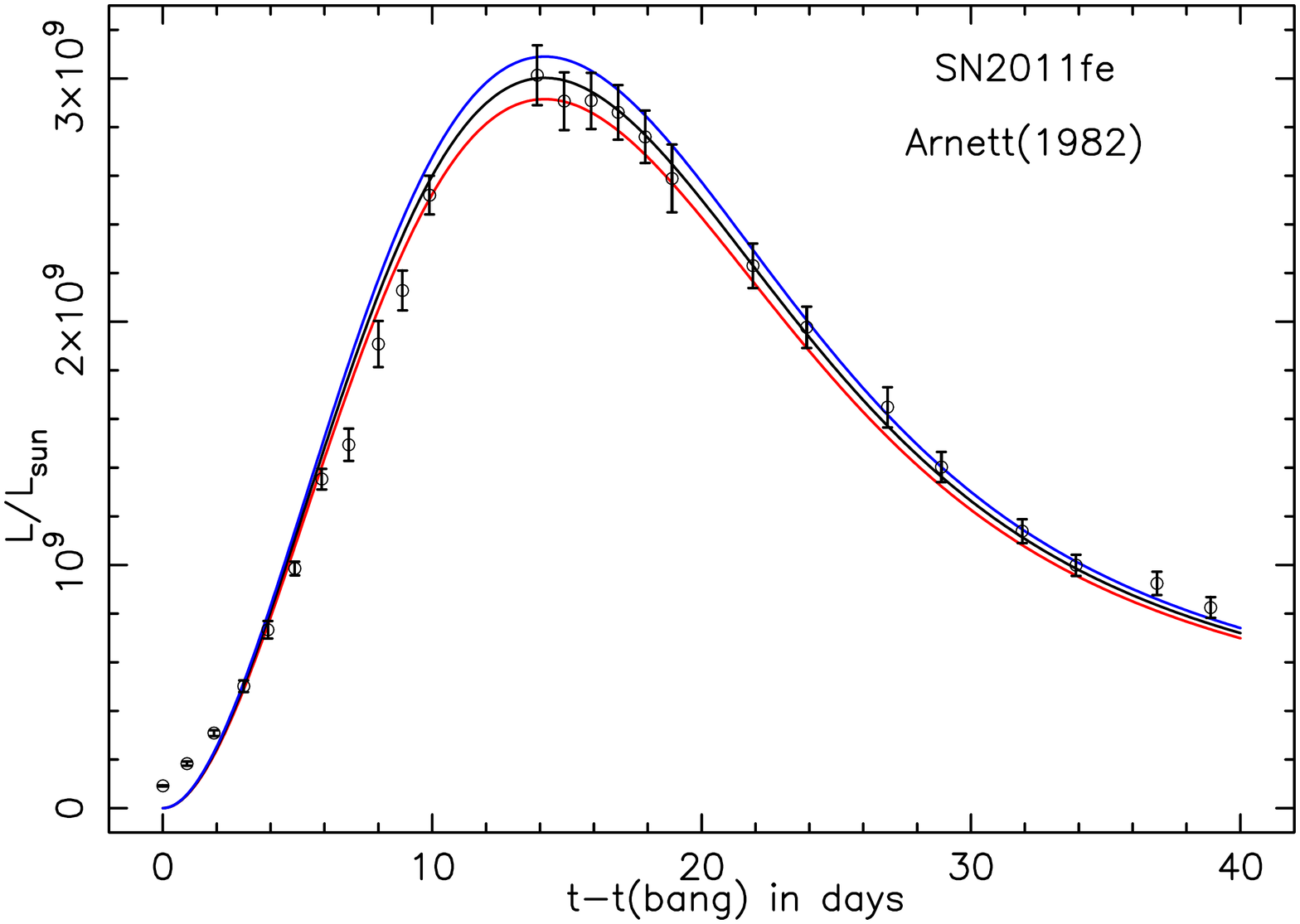}
\caption{Comparison of 
bolometric luminosity from analytic models \citep{a82a} and from SN2011fe \citep{sn2011fe}. The  analytic models plotted here 
allow for $\gamma$--ray escape, as did the original ones, by using the deposition estimate from Monte Carlo simulations of \cite{cpk80}.  The three 
curves are distinguished by a slightly different value for the mass of $\rm ^{56}Ni$, $M_{Ni} = 0.546 \pm 0.016 M_\odot$, which agrees well with that inferred by \cite{sn2011fe}; see text and Table~\ref{table1}. The peak luminosity increases with  $M_{Ni}$.}
\label{fig_sn2011fe} 
\end{figure}

Because the theoretical  models preceded the acquisition of high quality data by decades, and the physical parameters are nearly the same now as then (cf. Table~\ref{table1}), the models are a prediction  in this sense. %\footnote{Here we will give a critical analysis of the original theory,  and indicate new insights implied by the better data.}. 
It is easy to capture the observed behavior with these theoretical models.  Fig.~\ref{fig_sn2011fe} illustrates this observationally for SN2011fe \citep{sn2011fe} using \citep{a82a}.

\placefigure{1}

 We consider the original case of massive CO white dwarfs  igniting $\rm^{12}C$ under conditions of high electron degeneracy as being appropriate to SN2011fe and KSN2011B; see \cite{nugent2011} for observational support.
Burning $^{12}$C and $^{16}$O to nuclei lying in the range $^{28}$Si--$^{40}$Ca releases almost as much explosive energy as burning the same nuclei to $^{56}$Ni ($q_{SiCa} \sim q_{Ni} $), but the ashes (Si--Ca) are not radioactive, and do not provide late heating. We adjust the Ni production to fit the peak luminosity. We assume that comparable amounts of Si--Ca and Ni are produced, which determines the explosion energy. {\em We adjust the effective opacity to fit the shape of the light curve.}
The values chosen for SN2011fe light are documented in Table~\ref{table1}. 

\begin{deluxetable}{llll}
%\tablewidth{200pt}
\tablecaption{Base model for SN2011fe\label{table1}}
\tabletypesize{\small}

\tablehead{ 
 \colhead{Symbol}  &\colhead{Fig.~\ref{fig_sn2011fe}} &\colhead{A82$^a$} 
}
\startdata
  $M_{ej}$$^b$ ($M_\odot$) & $1.40$ & $1.45$ \\
 $M_{Ni}(0)$$^c$($M_\odot$) & $0.546\pm0.016$ & 0.5---0.7\\
 $R(0)$ & $4.0 \times 10^8 $ cm & 0---$10^{13}$\,cm \\
%Velocity scale & $v_{sc}$ & $1.0752E9$ cm s$^{-1}$ \\
 $\kappa_t$ ( cm$^2$ g$^{-1}$) & $0.09$ & 0.08 --- 0.1\\
 $E_{sn}$ ($10^{51} \rm \, erg$) &  1.2  & $\sim 1$  \\
 $y^2$ & $0.573 $ & 0.8 --- 1.2 \\ 
\enddata
\tablenotetext{a}{Values from \cite{a82a} for comparison.}
\tablenotetext{b}{Total stellar mass for SNe Ia.}
\tablenotetext{c}{Increase to $\sim 0.58 M_\odot$ for variable leakage \\
opacity (see text). }

\end{deluxetable}

\placetable{1}
           
With a mass of $\rm^{56}Ni$ of $0.546 M_\odot$,
after 14.14 days (but see \S\ref{section-early}) a maximum luminosity of $3.00 \times 10^9\,L_\odot$ is reached; which fits the data of \cite{sn2011fe}, who quote $3.04 \times 10^9\,L_\odot$. 
They  use $M_{Ni} = (0.44 \pm 0.08 \times (1.2/\alpha) M_\odot$, with $\alpha=1.2$ to infer the mass of $\rm^{56}Ni$; this expression is from \cite{gg12} and has assumptions regarding three-dimensional (3D) effects (see \cite{am16key} for a discussion of 3D and resolution issues). We prefer to take $\alpha=1$ so 
that the observationally inferred value is $M_{Ni} = 0.528 \pm 0.08 M_\odot$, which agrees better\footnote{Accounting for a variable leakage rate (see below) gives a longer rise to peak light, and  $M_{Ni} \sim 0.58 M_\odot$.} with the actual 3D simulations of \cite{roepke}, who found $M_{Ni} \sim 0.61 M_\odot$.

At peak only $83.8$ percent of the instantaneous radioactive energy release is being deposited in our model, while  the rest escapes as x-rays and gamma-rays rather than as thermalized radiation. The mass fractions of $\rm ^{56}Ni$ and $\rm ^{56}Co$ are $0.197$ and $0.741$ respectively.  Decays of $\rm^{55}Co$ and  $\rm^{57}Ni$ have a negligible effect of this part of the light curve.

\begin{deluxetable}{lll}
%\tablewidth{600pt}
\tablecaption{Some Initial$^a$ values \label{table2}}
\tabletypesize{\small}

\tablehead{ 
\colhead{Variable} & \colhead{Symbol}  &\colhead{Value} 
}
\startdata
Velocity scale & $v_{sc}$ & $9.28 \times 10^8$ cm s$^{-1}$ \\
Hydro time & $\tau_h$ & $0.43$ s \\
Diffusion time & $\tau_d$ & $1.52\times 10^{12}$ s \\
Central Density & $\rho(0,0)^b$ & $1.04 \times 10^{7}$ g cm$^{-3}$ \\
Central Temperature & $T(0,0)$ & $8.77 \times 10^8$ K  \\
Optical depth & $\tau_t(0)$ & $ 3.73 \times 10^{14}$  \\
%L factor & $\Lambda_{fak}$ & $4.75 \times 10^{43}$ erg  s$^{-1}$\\
%L factor & $\phi_{fak}$ & $1.92 \times 10^{37}$ erg s$^{-1}$\\
\enddata
\tablenotetext{a}{After shock emergence.}
\tablenotetext{b}{Uniform, for consistency with $\gamma$ escape.}
\end{deluxetable}
\placetable{3}

Table~\ref{table2} gives some initial values constructed from these parameters. The initial radius divided by the velocity scale gives an expansion time scale $\tau_h \sim 4 \times 10^8 / 10^9 \sim 0.4$~s. The diffusion time for photons is larger, $\tau_d \sim 1.5 \times 10^{12}$~s. This implies that an enormous expansion must occur before photons leak out readily, and that the temperature is determined by a balance between adiabatic cooling and radioactive heating. After the explosion, the Hugoniot-Rankine relations for a shock wave imply that the enthalpy and kinetic energy are comparable. Spherical expansion will reduce the internal energy by a factor $\tau_h/t$, so that within a minute the internal energy is reduced to $1.2 \times 10^{-4}$ of its post shock value. This energy is converted into kinetic energy by work done by pressure in expanding the matter, ${ 3\over 10}M_{ej}v_{sc}^2 \rightarrow E_{sn}$,
and we have an expanding Hubble flow. The energy released by radioactive decay equals the reduced thermal energy after about $40$\,s; after that the internal energy results from radioactivity.
``Early'' observations ($t >$ 1 hour) of SNe Ia tell us primarily about thermal, not hydrodynamic effects.

This implies that (after 1 hour) the spatial distribution of internal energy closely tracks that of radioactivity, because that internal energy is caused by radioactive heating. This resolves an issue in \cite{a82a},  Eq.~13, in which this was {\em assumed} without justification.

At much later times, $t \sim 10^{15} {\,\rm cm} / v_{sc} \sim 10^6$ s, maximum light occurs, with thermal photons, x-rays and $\gamma$-rays leaking rapidly.   

\subsection{Early time light curve}\label{section-early}
Figure~\ref{fig_sn2011fe} shows an excellent agreement of the original theory with observations.
However the new data offer the promise of new understanding: 
the first three data points are higher than the theoretical ones. In reality the opacity is not constant as assumed; at early times the temperature and density are rapidly changing, on a time scale $ t \sim R(t)/v_{sc}$, so variations in opacity are plausible.

\begin{figure}[h]
\figurenum{2}
\includegraphics*[angle=-90,scale=0.27]{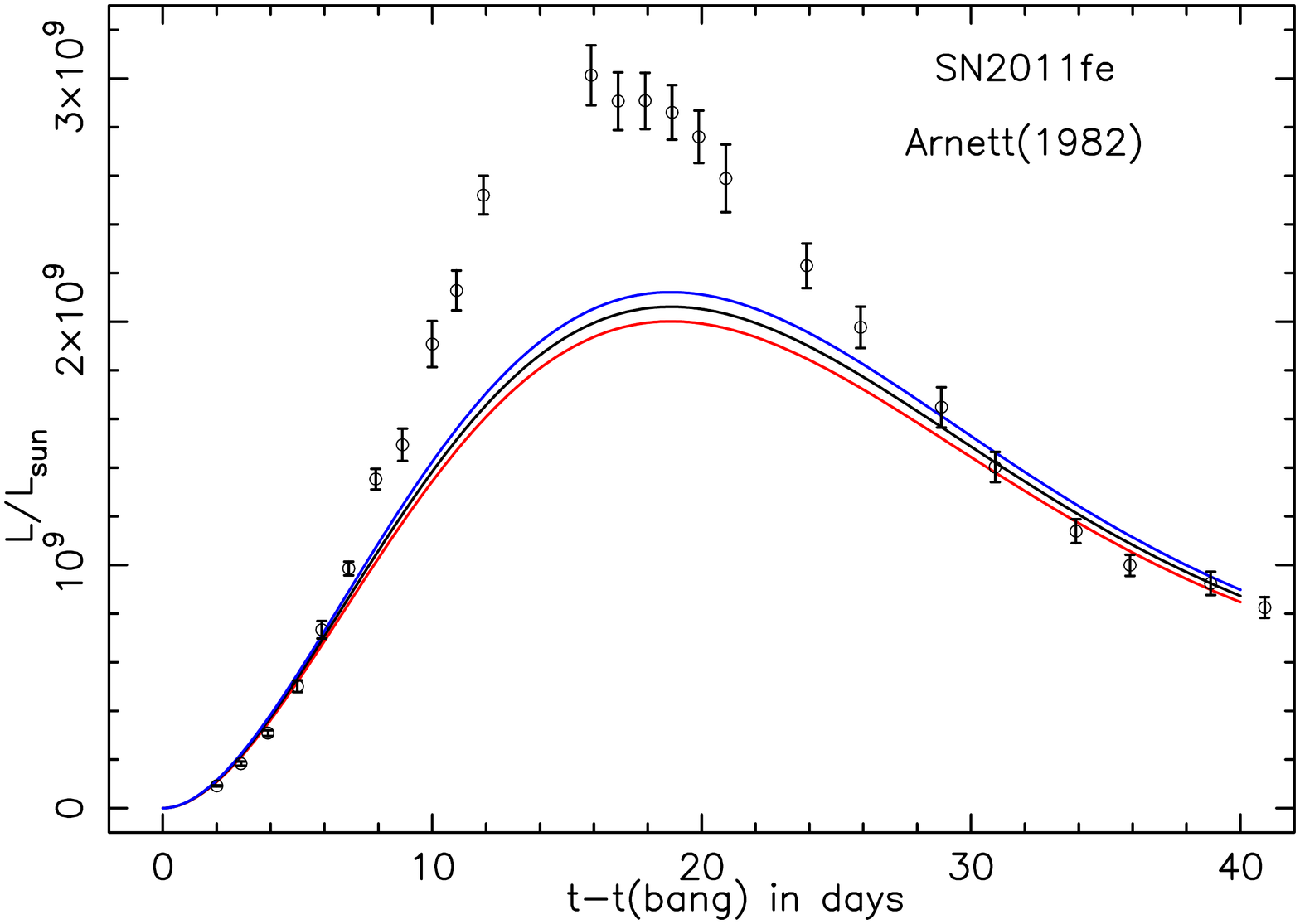}
\caption{Comparison 
bolometric luminosity from analytic models \citep{a82a} and SN2011fe \citep{sn2011fe}. Here the effective opacity is $\kappa_t = 0.2\, \rm cm^2 g^{-1}$, appropriate for early times when Thompson scattering dominates. The early points ($t < 6$ days) agree with the analysis of \cite{nugent2011}, but bang time is about 2 days earlier than in Fig.~\ref{fig_sn2011fe}. After about 8 days the analytic model has an effective opacity which is too large around peak light ($\kappa_t = 0.2 > 0.09\, \rm cm^2 g^{-1}$), but at 30 days after explosion, with gamma-escape important, the light curve again approaches the observed values again. 
}
\label{fig_sn2011fe-thom}
\end{figure}

Given Eq.~\ref{eq.Lxy}, and expanding $\Lambda(x,y)$ at early times  ($t/\tau_{Ni} = xy \ll 1$), we have
 $\Lambda(x,y) \rightarrow x^2 $ (Eq.~45 in \cite{a82a}).  The normalized luminosity is then
\begin{equation}
L/ \epsilon_{Ni} M_{Ni} =  t^2 /(2 y \tau_{Ni})^2, \label{eq-Learly}
\end{equation} 
which is the same form used by \cite{firth} but with $n=2$ and $a = 1/(2y\tau_{Ni})^2$. If $y^2 = ( \tau_d \tau_h / 2 \tau_{Ni}^2) \propto \kappa_t M_{ej} /v_{sc}$, is not strictly constant, but decreases  slowly with time\footnote{From $y^2 = 1.273$ in  Fig.~\ref{fig_sn2011fe-thom} to  $ y^2 = 0.573$ in Fig.~\ref{fig_sn2011fe}.}, $n>2$ results. Such a decrease in opacity is also needed to make the light curve in Fig.~\ref{fig_sn2011fe-thom} merge into that in Fig.~\ref{fig_sn2011fe} at peak luminosity.

This differs from the more limited ``blast wave'' approximation usually  used to support a $t^{-2}$ behavior: here the effective temperature $T_{eff}$ and the velocity at the photosphere $v_{photos}$ are not required to be constant. They are only constrained to radiate the luminosity leaked from inside. %It is the increase in the  surface of leakage that gives the $t^2$  dependence.

\cite{nugent2011} found the initial rise of SN2011fe to be well-described by a radiative diffusion wave with
an opacity due to Thomson scattering, $\kappa = \kappa_{th}$; Figure~\ref{fig_sn2011fe-thom} confirms this, using the parameter values of Table~\ref{table1}, but increasing $\kappa \rightarrow \kappa_{th} \approx 0.2\,\rm cm^2\,g^{-1}$. Now the earliest three points lie on the theoretical curve, but the luminosity in the peak is too low because the opacity is too high; %(gamma and x-ray transport becomes important, so
 $\kappa \sim 0.09\rm\, cm^2\, g^{-1}$ at peak is needed to model the escape of thermal energy. At later times ($t > 30$ days after explosion) the shape of the light curve is controlled by increasing gamma-ray escape, and the numerical value of the effective opacity for thermal radiation is not so important.
 
\placefigure{2}

The ``bang time'' (time of explosion) 
in Figure~\ref{fig_sn2011fe-thom} is now earlier by 2 days. \cite{firth} found a rise time of $18.98\pm0.54$ days for 18 SNe Ia, ranging from $15.98$ to $24.7$ days. For SN2011fe
 our value of $14.14 + 2 \sim 16.1$ days agrees well with the \cite{sn2011fe} value of $16.58\pm 0.14$ days.

 \cite{firth}  %fitted their early data to a form $L = a t^n$ for a normalized flux $L$, with $t$ measured from the estimated onset of explosion (bang time),  and 
 find  the exponent of the power law fit to the early rise to be $n=2.44 \pm 0.13$. Although Figure~\ref{fig_sn2011fe-thom} begins as $L \propto t^2$, merging upward to  Figure~\ref{fig_sn2011fe} requires an steeper increase in the transition and thus $n>2$ if a power-law fit is made for constant $a$  (i.e., $L\approx a\, t^n$). 
 
 We note that this behavior of $n$ may
be a general property of the {\em decoupling of super-thermal photons} with continuing expansion 
rather than the {\em distribution} of $\rm^{56}Ni$, as suggested by \cite{piro}, or a combination of the two.
It remains to disentangle these effects; they both deal with the effects of radioactive decay heating.

\subsection{Opacity}\label{section-opacity}
Consider a simple leaky-bucket model for radiation loss\footnote{This discussion was heavily influenced by \cite{pe00a,pe00b}.}; frequency regions of low opacity are ``holes'' and determine the leakage rate. The highest energy photons (gamma and x-rays) have their longest mean-free paths for scattering, and these interactions fill the holes because of their relatively weak frequency dependence (see \S\ref{S-holes}). At early times the gamma and x-rays are downscattered into the thermal range, so the the relevant scattering cross section for filling the holes is that for Thomson scattering, and $\kappa_{th} = 0.20\rm\,cm^2\,g^{-1}$. 
At late times a floor on the opacity is derived from the Klein-Nishina transport cross section to be $\kappa_{KN} = 0.067\rm \, cm^2\, g^{-1} = \kappa_{th}/3$. Fitting the shape of the peak in Figure~\ref{fig_sn2011fe} gave us $\kappa_{t} \approx 0.09\rm\,cm^2\,g^{-1}$ which lies between these limits.  
Apparently the effective opacities for energy leakage in SNe Ia are (1) small, (2) vary little, ($0.067 < \kappa_t < 0.2$), and (3) are insensitive to composition, a surprising result given the complexity  of the spectra and the problem. This is consistent with the discussion in \cite{dessart14}, \S3.2.
It seems that {\em leakage of energy} rather than {\em line formation} is the key issue for bolometric
light curves; spectra are the opposite.

It appears that a modest and plausible variation in the effective opacity (escape probability) will allow an excellent description of the SN2011fe light curve, e.g., something like Figure~\ref{fig_sn2011fe} with Figure~\ref{fig_sn2011fe-thom} for the first few days would result from a slowly varying (almost constant)  opacity. 

\section{Opacity and Energy Leakage} \label{leakage}

 In Section~\ref{feLightcurves} we found that the effective  opacity to be the remaining parameter, from the original five, which needed to be used to adjust the light curves to fit the data from SN2011fe.
The analytic solutions for the bolometric light curve require a rate at which photon energy, both thermal and non-thermal,  leaks from the ejected mass; this may be quantified by a ``leakage time''. 

\subsection{Leakage time}
The thermal leakage is estimated by a radiative diffusion model, which involves the integrated effect of the opacity over the structure and time \citep{a82a,pe00a,pe00b,pe00c}. Although the local opacity will vary sensitively with the local density, temperature and photon frequency, the leakage time integrates these effects over a dynamic structure \citep{pe00b}, giving a smoother variation in global properties. 
The term ``opacity'' in the analytic solutions is a placeholder for a leakage time. This ``effective opacity'' was taken to be a constant\footnote{\cite{pe00b} estimated $\kappa \approx 0.1\rm\,cm^2\,g^{-1}$, which is almost identical to our value inferred from the SN2011fe light curve.} 
(for analytic simplicity) which approximates the same leakage of radiative energy as a physically correct simulation would (in practice, this may only mean that the choice of effective opacity fits the observations). 

As seen above, this should include effects due to x-ray and $\gamma$-ray {\em transport}, not just {\em deposition} as in \cite{cpk80}. As maximum light is approached, energy transport by superthermal photons (x-rays and $\gamma$-rays) becomes comparable to that by ``thermal photons'' (see \cite{pe00b} for discussion of thermalization).  As maximum light is approached non-thermal heating becomes less localized and mean-free-paths become longer. If we regard this as a second channel for energy flow, the effective opacity would be (roughly) the inverse average over both thermal  and non-thermal channels, with the lower opacity channel carrying more energy.
%$1/\kappa_t \sim 1/\kappa_{non} + 1/\kappa_{thermal}$, so it falls below the purely thermal value, as we saw for SN2011fe.

\subsection{Conditions at maximum light}

Table~\ref{table3} shows the values of selected variables at maximum light for the models presented in Figure~\ref{fig_sn2011fe}.  The temperatures lie in the range $ 1 \times 10^4$ K to $2 \times 10^4$ K, which are commonly found in normal stars. 
Due to the large radiative conductivity, the SN  has a relatively shallow temperature  gradient.

Unlike the temperatures, the densities (of order $3\times 10^{-13}$ g cm$^{-3}$) are considerably lower than encountered in stellar atmospheres.  
These low densities suggest the question: how valid is the assumption of collisional equilibrium?
As discussed in \cite{pe00b},  thermal velocity $v \ll c $ implies that collisional equilibrium is difficult to attain in supernovae,  and thermal equilibrium comes from radiative interactions. 

\begin{deluxetable}{lll}
%\tablewidth{600pt}
\tablecaption{Variables at Peak Light$^a\label{table3}$}
\tabletypesize{\small}

\tablehead{ 
\colhead{variable} & \colhead{value}
}
\startdata
$t_{max}-t_{bang}$ & 14.14 days$^b$  \\
$L_{max}$ & $3.00 \times 10^9\, L_\odot$ \\
m(bol) & -18.97 \\
 $\rho_c$& $4.56 \times 10^{-13} \,\rm g \, cm^{-3} $ \\
 $T_c$ & $1.92 \times 10^4 \rm\,K $ \\
 $T_{eff}$ & $1.06 \times 10^4 \rm\,K $ \\
 $\tau_{h\nu}$(thermal) & 46.6 \\
 $\tau_\gamma$(superthermal) & 15.5 \\
 D($\gamma$-deposit) & $0.838$\\
 $^{56}$Ni & $0.199$ \\
 $^{56}$Co & $0.740$ \\
 $^{57}$Ni & $3.12 \times 10^{-5}$ \\
 $^{57}$Co & $0.0237$ \\
\enddata
\tablenotetext{a}{See Figure~\ref{fig_sn2011fe}.}
\tablenotetext{b}{Add $\sim2$ days to account for extension seen in Figure~\ref{fig_sn2011fe-thom}.}
\end{deluxetable}
\placetable{3}

\subsection{LTE Fe opacities and holes}\label{S-holes}
\begin{figure}[h]
\figurenum{3}
\includegraphics*[angle=0,scale=0.36]{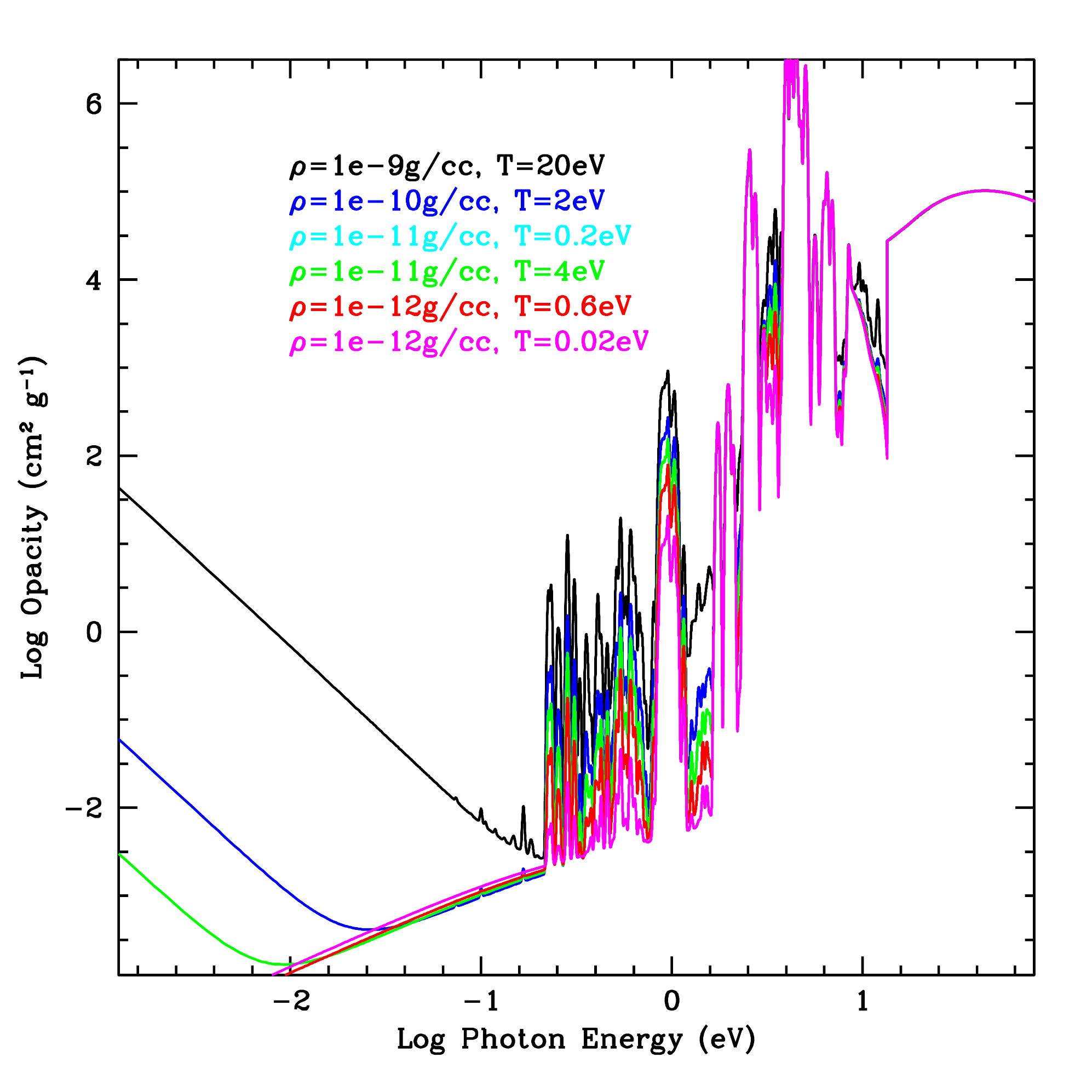}
\caption{Logarithm of Fe opacity as a function of photon energy for various combinations of density and temperature using an 
extension of the OPLIB database \citep{magee95} (courtesy Chris Fontes).  Note that as the ejecta expands (and the density and temperature decrease), the opacity in a given band can both decrease and increase. At low density ($\rho = 10^{-12}\rm\,g\,cm^{-3}$) gaps open in the atomic opacity, which decreases toward the IR frequencies.
}
\label{feopacity}
\end{figure}

Chris Fontes has calculated new LTE opacities for these low densities, which are shown in Figure~\ref{feopacity}. 
These opacities are produced by the 
OPLIB database team \citep{magee95} with new calculations for the low densities needed for light-curve calculations, and
shown in Fig.~\ref{feopacity}.
Following \cite{pe00b} we accept that LTE is a flawed approximation in detail, but use it to suggest qualitative behavior.
In a simplified leakage scheme for transport of energy, we assume a single opacity
for the transport, but the opacities vary dramatically with photon
energy.  
For leakage schemes, the transport is dominated by the dips in the
opacity, not the peaks.  The energy where this dip occurs depends both
on the temperature and density of the supernova ejecta, and the leakage
opacity may vary both up and down as the ejecta expand, ranging
within the visible bands from below $0.001\rm\,cm^2\, g^{-1}$ to above
$10\,\rm\,cm^2\, g^{-1}$.  

\placefigure{3}

As the densities approach supernova values ($\rho < 10^{-12}\rm\,g\,cm^{-3}$),  the opacity in Fig.~\ref{feopacity} shows gaps with values $\log_{10} \kappa \sim -2$, around and below the LTE temperature of 1eV. Guided by \cite{pe00b} we assume that this property is set by atomic structure and will be qualitatively true for non-LTE distributions of states; see also \cite{dessart14,dessart,dessart16}, who find that their spectra are surprisingly insensitive to the opacity controling the radiation flow. 

Non-thermal radiation may affect the degree of ionization around peak light, when thermalization is weakening, and affect the leakage and transport of energy. Such effects have been subsumed in the ''effective opacity''.

This appears to support the inference from \S\ref{feLightcurves} that the effective opacity may be controlled by scattering processes,   Thomson and Compton, which close leaks.

\subsection{Leakage and transport}
\begin{figure}[h]
\figurenum{4}
\includegraphics*[angle=0,scale=0.36]{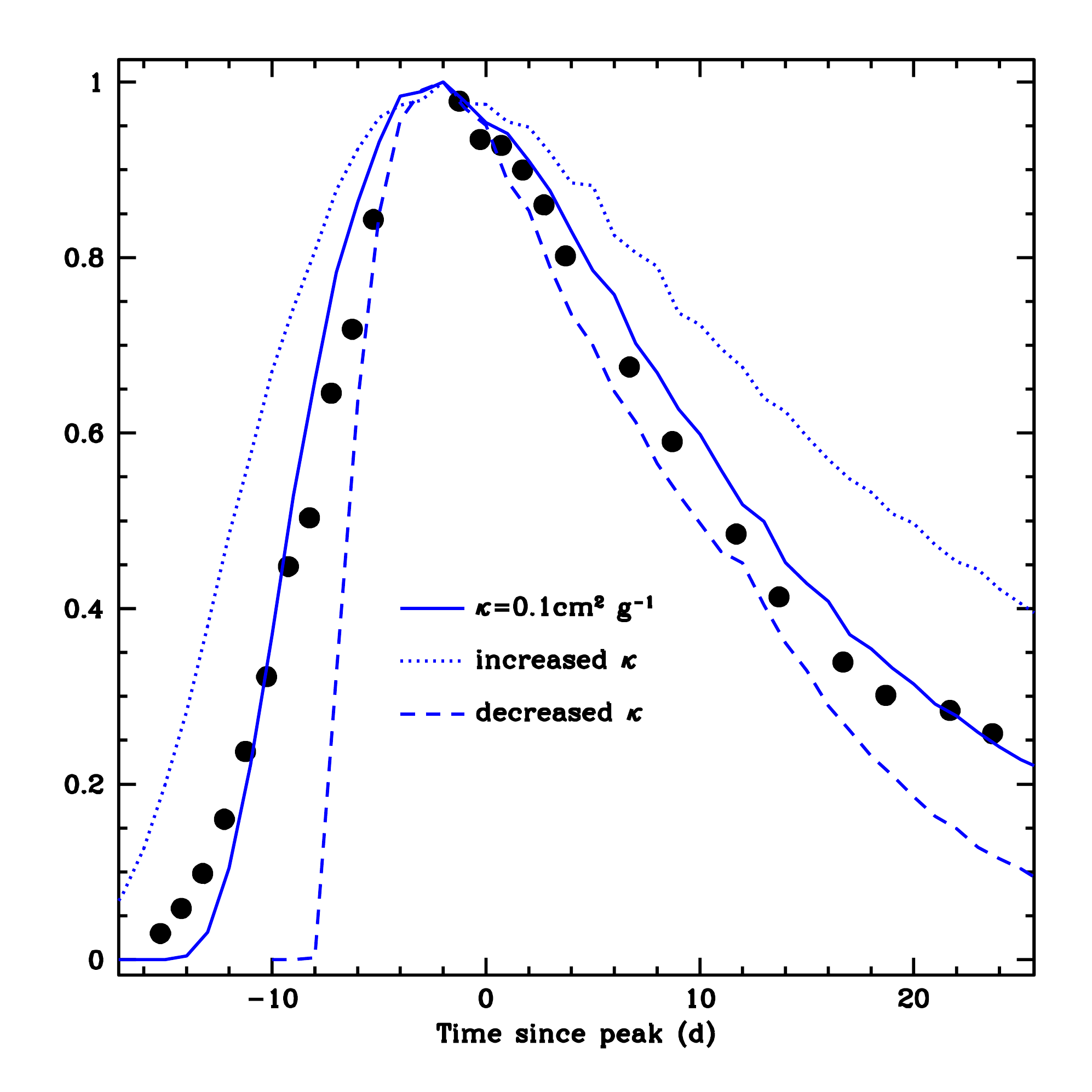}
\caption{Comparison of bolometric luminosity from SN2011fe
  \citep{sn2011fe} (solid dots) using our diffusion code with three opacity
  prescriptions: (a) constant $\kappa=0.1\rm\,cm^2\,g^{-1}$ (solid),
 (b) increase at low temperatures: $\kappa$ increases to $1\rm\,cm^2\,g^{-1}$
  below 1\,eV (dotted), and (c) decrease at low temperatures: $\kappa$
  decreases to $0.01\rm\,cm^2\,g^{-1}$ below 1\,eV (dashed).  Increasing the
  opacity broadens the light-curve, decreasing narrows it.
 These explosions assume an explosion mass of $1.4\,{\rm M}_\odot$, an 
explosion energy of $1.3\times10^{51}\,{\rm
    erg}$ and a nickel mass of $0.6\,$M$_\odot$}
\label{opacsim}
\end{figure}

To better understand the effect of such opacities, we have
implemented an opacity switch in the simplified transport
code of \cite{bayless} that allows the opacity to move up or down by an
order of magnitude as the temperature drops below 1\,eV ($1.16\times
10^4$\,K).  Figure~\ref{opacsim} shows the light curves from a standard 
model assuming $\kappa=0.1$ as well as the opacity switch that increases 
and decreases the opacity.  These light-curves demonstrate the sensitivity to opacity.  With a full transport solution, we 
might expect opacity variation to produce observable features in the light curve (which we do see), 
but the basic trends with the slopes will not change dramatically.
Notice that  for the increased opacity case, the lightcurve begins to rise earlier, as in Figure~\ref{fig_sn2011fe-thom}.

\placefigure{4}

Until these issues are  convincingly resolved, the leakage time should be considered an adjustable parameter, strongly constrained by observation, 
constrained less strongly by microscopic theory, but  not yet one solidly founded in experiment and simulation.

\subsection{Filaments and mixing}\label{filaments}

 In addition to issues associated with line formation and leakage,
there are the difficulties associated with non--spherical variation inevitably resulting from instabilities in the explosion.  Theoretical models of supernovae which are spherically symmetric have no angular resolution to deal with this broken symmetry. Mixing is treated in a ad-hoc manner. Observed young supernova remnants have a pronounced filamentary structure, a heterogeneity which may have developed during this epoch of instability.

How might this happen?  We summarize the discussion in  \cite{am16key} and \cite{321D}. Explosions are rife with instabilities. Most notable are
Rayleigh-Taylor and Richtmeyer-Meshkov instabilities, which also appear in simulations of convection by bottom heating.
Heat gives rise to plumes which develop a mushroom cloud shape. In a convective region the upward motion is bounded and turned back. Because of the astronomically large Reynolds number, the flow is turbulent and gives rapid mixing, which in high-resolution simulations is essentially complete in two turn-over times (four transit times). Suppose the heating is violent, and the motion is not contained (an explosion). Then only one transit time is available for mixing, which is global yet incomplete. Turbulence is frozen by expansion. A fundamental feature of turbulence is high vorticity: vortex filaments are the ``sinews'' of turbulence \citep{llfm,pope}.
Further expansion of the supernova may lead to a regime of thermal instability, in which denser regions radiate, cool, and are further compressed, leaving the sort of filamentary structure seen in young supernova remnants.
Is it possible to test this idea with observations of supernova light curves? The Kepler observations of supernovae  may bear on this question.

\section{Supernova KSN2011B}\label{section-ksn}

\begin{figure}[h]
\figurenum{5}
\includegraphics*[angle=0,scale=0.27]{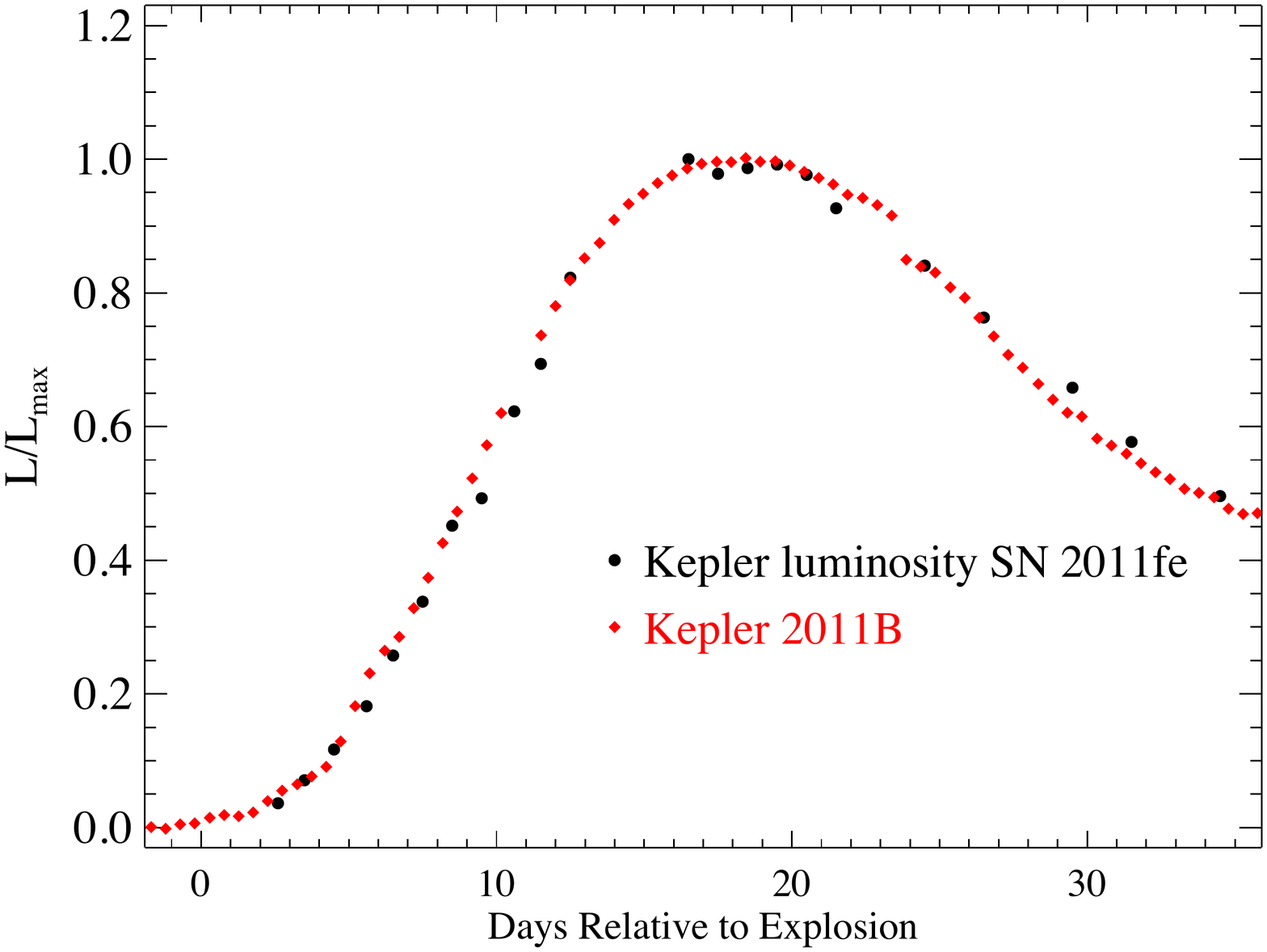}
\caption{  Comparison of  Kepler-normalized luminosities (i.e., in
  the Kepler filter bandpass) for  SN2011fe and KSN2011b
  \citep{keplersn}. When normalized at peak light the two SNe Ia look
  very similar.  The time axis was translated to match the point
  of explosion.  We took the time of explosion for SN~2011fe from
  \cite{sn2011fe} (2.6 days before the first spectrum) and from
  \cite{keplersn} for Kepler 2011b (18.1 days before maximum).}
\label{kepler1}
\end{figure}
\placefigure{5}

While the Kepler satellite does not have  wavelength
coverage comparable to that which became available for SN2011fe, the Kepler
coverage is broadband, has high cadence in time, has small errors from statistical fluctuations, 
and the light curves track the data from SN2011fe well; see Fig.~1 in \cite{keplersn}. 
We suggest that the Kepler data provide a constraint on mixing in SNe Ia that is unattainable even with the high quality 
UVOIR data available for SN2011fe.

The spectra of SN~2011fe in \cite{sn2011fe} were obtained using the
SuperNova Integral Field Spectrograph \citep{Lantz04} on the 88''
telescope on Mauna Kea and are presented as fully calibrated.  We multiplied the
spectra by the filter function for Kepler (a wide filter, $\sim 4400
- 8800$
A)\footnote{http://keplerscience.arc.nasa.gov/the-kepler-space-telescope.html}.
The Kepler system is not fully calibrated, so there is no measured
zero point to place Kepler magnitudes onto a known scale.  We used a zero
point similar to that for standard filters, a procedure which gives a reasonable answer, but 
still contains some arbitrary scaling. Even evaluating the flux
directly from the spectra through the filter requires some knowledge
of the calibration of the filter.  We can construct a Kepler magnitude
and a flux in the Kepler filter for each spectrum that are all
consistent relative to each other, but there is still an uncertain
absolute scaling. Actual bolometric corrections are hard to do without
calibrating the Kepler system.

It appears that SN2011fe and KSN2011b are similar enough (Fig.~\ref{kepler1})
that we may use this similarity in the data to establish a
correspondence between the bolometric scale of SN2011fe and the
Kepler supernovae.

\subsection{A Calibration}
\begin{figure}[h]
\figurenum{6}
\includegraphics*[angle=0,scale=0.27]{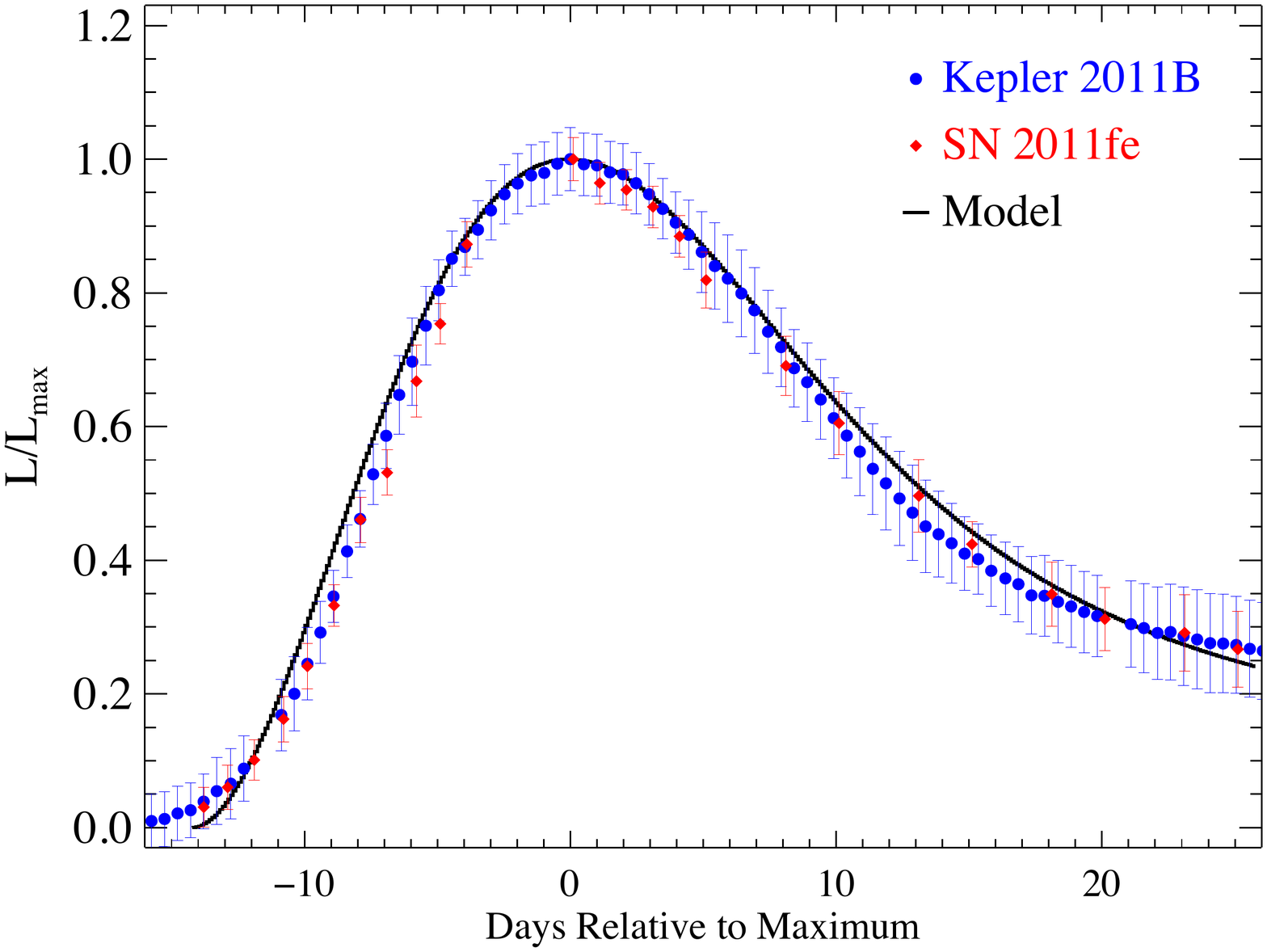}
\caption{Comparison of the bolometric luminosity of SN~2011fe
  \citep{sn2011fe} and adjusted data (see text) based on KSN2011b
  \citep{keplersn}.  The light curves presented by \cite{sn2011fe} are
  also obtained by convolving filter functions with their spectra.
  They estimate the errors in their magnitudes using the flux
  calibration procedure for their spectra and we adopt the same
  1$\sigma$ errors per epoch.  The errors for KSN2011b incorporate
  those errors as the bolometric correction is applied.  The choice of
  explosion time (the precise zero of the x-axis) depends upon the
  algorithm used, and may be uncertain by a day or so (see
  \S\ref{section-early}). }
\label{kepler0}
\end{figure}

%As the Kepler-magnitude light curve of SN~2011fe is so similar to KSN2011b (Fig.~\ref{kepler1}), 
From Fig.~\ref{kepler1}, 
we infer that the bolometric light curve of KSN2011b
would be similar to SN2011fe. There are many caveats that need to be
considered for such an inference, such as the fact that SN2011fe and
KSN2011b may have different bolometric characteristics.  Given the
striking similarity of the two light curves in the Kepler band,
though, we believe that this is a plausible inference.  Using the
bolometric luminosties from \cite{sn2011fe}, we establish a
`bolometric correction' for our derived Kepler-magnitude light curve
of SN 2011fe.  We then apply the same corrections to KSN2011b to
derive a bolometric light curve.  We take the
flux and treat it as a relative luminosity (L/L$_{\rm max}$); this
ratio also removes concerns about absolute scaling.
Fig.~\ref{kepler0} shows the time evolution of the bolometric L/L$_{\rm max}$ for
both SN2011fe and KSN2011b.

\placefigure{6}

Now that we have a calibration, how does the  best Kepler supernova compare
with the analytic curves?  %The best example, 
KSN2011b, from
\cite{keplersn} is shown in Fig.~\ref{kepler0}.   Random
  photometric errors of KSN2011b are small.  The uncertainties in the
  SN 2011fe bolometric luminosities and the bolometric corrections
  applied to the KSN2011b Kepler magnitudes dominate the error bars. There is evidence for a well-defined and
smooth light curve, but systematic biases might affect details of shape fitting. 

Again we see a
deviation between observations and analytic light curves at very early
times, which we attribute to use of a constant effective opacity. 
After this first deviation, the curves are strikingly similar, as the
higher cadence of the Kepler data makes clear. Although the observed
supernovae might not be identical, so that fitting with different
theoretical parameters would be legitimate in principle, we have used exactly the
same theoretical parameters as in Fig.~\ref{fig_sn2011fe} in order to
illustrate just how similar these light curves seem to be.

{\em Because of the steady, fast cadence of observations, KSN2011b provides unique new information on SNe Ia: the smoothness of the light curve.} While KSN2011b tracks SN2011fe within its error bars, the KSN2011b light curve is noticeably smoother. There is no indication of any bumps due to collision of ejecta with circum-supernova matter, as with type Ibn and IIn events. 
The smoothness of 1D theoretical light curves result from ad-hoc ``box-car'' mixing; simulations without such mixing are not smooth \citep{pe00a,dessart,dessart16}.
Explosions are unstable to 3D mixing which can leave an imprint on the young supernova remnant as well as the light curve \citep{am16key}.
SN2011fe does exhibit larger fluctuations in its light curve, consistent with its observational error bars.

\section{Arnett's rule}\label{rule}

It has been shown, for one-zone models  \citep{a79},  with constant opacity and radioactive heating by $^{56}$Ni and  $^{56}$Co decay, that
\begin{equation}
 L_{peak}  /M_{ej} \approx \epsilon_{NiCo}(t_{peak}) . \label{alaw1}
 \end{equation}
 where $ L_{peak}$ is the luminosity  at maximum light, $M_{ej}$ is the ejected mass, and  $\epsilon_{NiCo}(t_{peak})$ is the instantaneous rate of energy production from radioactivity at that time.
From a solution by separation of variables (radius and time), a more accurate expression was derived,
\begin{equation}
 L_{peak}  /M_{ej} = \epsilon_{NiCo}(t_{peak}) D , \label{alaw2}
 \end{equation}
 where $D$ is the deposition function of \cite{cpk80}, which is a function of ``optical" depth for $\gamma$-rays  (see \S\,IV.a in \cite{a82a}). For SN2011fe, $D = 0.838$ at  maximum light (Fig.~\ref{fig_sn2011fe}), which corresponds to a correction to Eq.~\ref{alaw1} of $\approx 20$ percent (increase) in estimated radioactive mass.
 
\subsection{A Derivation}\label{section-derive}

 Here we present a new derivation which is more concise, general, and hopefully clearer. We emphasize that, as in \cite{a82a}, it is the {\em instantaneous total heating rate of the plasma} which is constrained, not the radioactive decay rate, which may explain at least  part of the discrepancy with \cite{dessart,dessart16}. In addition we add a brief appendix summarizing a modern approach to solving for \cite{a82a} light curves, so that errors in implementation may be easily corrected.
 
The luminosity due to radiative diffusion may be written as
\begin{equation}
L =   a T^4 R^3 / \tau_{dif}, \label{eqL}
\end{equation}
where
\begin{equation}
\tau_{dif} \sim  \kappa M / \beta cR,
\end{equation}
and $\beta \approx 13.7 $ is a dimensionless form factor from integration of the diffusion equation over space. It includes the geometric factor for spherical geometry, and is very slowly varying with mass-density structure; see  \cite{a80}, Table~2 and \S VI.

At any extremum in $L$, either a peak or a dip, $dL/dt=0$.
The time derivative of Eq.~\ref{eqL} is 
\begin{eqnarray}
d\ln L/dt &=&  4d\ln T/dt + 4d\ln R/dt \nonumber\\
&-& d\ln\kappa/dt. \label{eqdLdt} \label{eq_LTdS}
\end{eqnarray}
For a slowly-varying escape time, the term $d\ln\kappa/dt$ will be small; see discussion in \S\ref{leakage}.
This is also consistent with the good fits to the light-curve data shown in  Fig.~\ref{fig_sn2011fe}.

A second equation involving the luminosity is first law of thermodynamics, which may be written {\em after spatial integration} as
\begin{eqnarray}
 L/ M \sim  \epsilon (t) D
- dE/dt - PdV/dt  .
\end{eqnarray}
For homologous expansion (Hubble flow) $$d\ln V/dt = 3 d\ln R/dt$$
and for a radiation-dominated plasma, $$E=3PV=aT^4V,$$ so
\begin{equation}
dE/dt + P dV/dt = E( 4d\ln T/dt  + 4d\ln R/dt). \label{eq_TdS}
\end{equation}
This requires a 3D simulation of high resolution to compute accurately \citep{am16key}, but deviations from Hubble flow may be thought of as compressional heating and expansional cooling from an inelastic collision within the supernova ejecta  (\cite{a80}, \S\,VIII), or with surrounding matter \citep{nathan}. 
Using Eq.~\ref{eq_LTdS} and \ref{eq_TdS} we have
\begin{eqnarray}
 L/ M \sim  \epsilon (t) D
- E( d \ln L/dt + d \ln \kappa/dt)  .
\end{eqnarray}
At any extremum in $L$, $dL/dt=0$, so at that time we have 
\begin{equation} 
L /M = \epsilon D - E\, d\ln\kappa/dt , \label{eq.rule}
\end{equation} 
for {\em any plasma heating rate $\epsilon D$.
This will be true at multiple peaks and at  dips as well.} This is a more general form of Eq.~\ref{alaw2}.

As emphasized in \cite{a82a} SNe Ia are special in that the kinetic energy source and synthesis of radioactivity are more directly linked than in SNe Ibc.
Heating  from other causes would have the same qualitative effects as radioactive heating. The $\epsilon D$ in Eq.~\ref{eq.rule}
may be generalized to be the {\em sum of all processes heating the ejecta.} It may be affected by relativistic jets (the fraction of their energy which is thermalized), magnetars, pulsars, fluid accretion (fall-back) onto a neutron star, and so on,  as well as radioactivity. SNe Ia models are simple in that they have only one heating source and no collapsed remnant.

In Fig.~\ref{fig_sn2011fe} there are two extrema in luminosity: at the time of explosion (bang time) and at maximum light. The first peak at very early time is due to shock breakout, and occurs at such small radii that it has negligible effect on the luminosity for SNe Ia. Eq.~\ref{eq-Learly} gives the ``initial'' luminosity after radiative equilibrium and Hubble flow have been established (see \S\ref{feLightcurves}). 

For SNe Ibc, the initial radius is not so small.
The first peak in SN2008D (Fig.~\ref{SN2008D}) may have a contribution from compressional heating as the supernova shock break-out occurs, as well as from radioactivity in the outer, fast ejecta. SNe Ibc may be showing 
{\bf evidence in light curves for fluid dynamic heating/cooling in non-homologous flow that is comparable to the radioactive heating rate at peak. }
%{\color{red} This may be the case for the simulations of \cite{dessart16}. 
%{\color{red} quote some numbers; reflected shock from core/mantle}
Multiple-peaked light curves like SN2016gkg \citep{tartaglia} might also be interpreted in this way.

 \subsection{A Thought Experiment}
 Consider a thought experiment which has observational implications: suppose a plume of $^{56}$Ni rapidly expands so that all $\gamma$-rays escape. This energy is lost, and does not heat the plasma or contribute to the light curve, except later when kinetic energy from the positron channel of $^{56}$Co decay becomes significant. {\em There would be an inconsistency in the amount of radioactive mass estimated from the pre-nebular light curve and the radioactive tail. }
 In a less extreme case, we note that 
 at early times $\gamma$-rays are always trapped, and only escape freely as expansion makes this possible.
 There will be an enhancement of the light curve {\em before} the $\gamma$-rays easily escape, and possibly an early peak in the light curve before the dominant one. Such a plume might be expected to have high entropy, so that the $^{56}$Ni might be formed by the ``$\alpha$-rich freezeout'', possibly producing detectable amounts of radioactive $^{44}$Ti.
 This is an example of how 3D mixing can deviate qualitatively from 1D, and why Eq.~\ref{alaw2} is to be preferred over  Eq.~\ref{alaw1}. See \S\ref{sn2008D} below.
 
 \subsection{SNe Ia}
 \cite{sblondin} found agreement to within 10 percent with ``Arnett's rule'' (apparently  Eq.~\ref{alaw1}) for their set of delayed-detonation models of SNe Ia, using the CFMGEN code.
 In  \S\ref{section-ksn} we showned that the light curve for the type Ia supernova, KSN2011b, was a smooth curve. The bolometric light curve directly measures the rate of  leakage of thermal energy from the supernova (see \S\ref{leakage}), and because the curve is smooth, this leakage may be approximated by a slowly varying average opacity. 
This contrasts to  the complex variation of the local opacity seen in \S\ref{S-holes} and CFMGEN.  Evidently integration over frequency, ionization, and mean-free-path  tames the unruly local  behavior of the opacity to give a better behaved leakage rate (as averaging  over the Kolmogorov cascade does for turbulence). 

We saw in  \S\ref{section-early}, however, that the leakage opacity  is not strictly constant, just slowly varying, and gives an increase in
$\Delta_{rise} = t_{max}-t_{bang}$ of 2 days, upon integration of the SNe 2011fe (Ia) light curve. This brings
$\Delta_{rise}$ from 14.14 days to 16.1 which agrees well with \cite{sn2011fe} who observe $16.58\pm0.14$ for SN 2011fe.
Near maximum light,  Eq.~\ref{eq.rule} may be written as
\begin{equation} 
L /M\epsilon D \approx 1 -  (E/\epsilon )d\ln\kappa/dt ,
\end{equation} 
which implies that opacity decreasing in time increases $\Delta_{rise}$ by shifting the peak to later time.

\subsection{SNe Ibc}
 In contrast,
\cite{dessart,dessart16} found errors of 50 percent using Eq.~\ref{alaw1}  for models of SNe Ibc; these errors they attributed to  the use of a constant mean opacity. However their conclusions may be weakened, not because of doubt about the CFMGEN results, but because their approximation to \cite{a82a} deviates from the original; they use 100 percent trapping of gamma-rays ($D=1$), and the integral approach of \cite{kkd13}.  \cite{dessart,dessart16} also state that they also must add a new $^{56}$Co decay from \cite{valenti} (which they correct further), rather than the original in \S\,VI.a of \cite{a82a}.
 As this rule of \cite{kkd13} and Eq.~\ref{alaw2} are both based on the first law of thermodynamics, it may be suspected that deviations between the two are most likely due, at least in part, to the equations being incorrectly applied, or to using different physics (e.g., $D=1$). 
%  The latter is mathematically equivalent to the energy conservation equation used by \cite{a82a}; s
See Appendix~\ref{appendixb} for a simpler mathematical formulation of \cite{a82a}.

\section{SN2008D}\label{sn2008D}

The similarity in light curves, of thermonuclear and core-collapse supernovae suggests that we attempt to use the procedures  for SNe Ibc as well.
%{\bf \cite{sblondin} have confirmed that the light curves of SNe Ia are well reproduced for their numerical models, but  \cite{dessart16} were less successful for SNe Ibc for reasons discussed in \S\ref{rule}.}
The approach of \cite{a82a} gives an opportunity to separate radioactive from non-radioactive heating in the light curves of SNe Ibc.
To illustrate this point we examine the light curve of the well observed SNe Ib 2008D.

Supernova 2008D was discovered at the time of explosion, which coincided with an X-ray event due to the break-out of the supernova shock wave \citep{soderberg}.  This was the first successful observation of the previously predicted break-out of a supernova shock \citep{colgate,wda71,cheval}, and to be associated with a bare core, or stripped-envelope supernova \citep{a77Tx8}. \cite{modjaz} presented a detailed discussion of the observational data (optical, infra-red, and X-ray) of this event, and \cite{bianco} summarized the results for 64 similar events, including this one. We base our discussion on the data sets in these papers, which contain a more detailed history and references.

SN2008D occurred in NGC 2770 ($D=31\pm2$ Mpc) and had a peak absolute magnitude of $M_V= -17.0\pm0.3$ mag and $M_B=-16.3\pm0.4$ mag \citep{modjaz}. The rise time in the V-band was $16.8\pm0.4$ days and in the B-band $18.3\pm0.5$ days, near the long end of the range for SNIbc. The observation of the X-ray outburst determines a more precise value for the explosion time, which is of value in fitting the data to theoretical models (see Section~\ref{section-early}). Integrating the observed bands, or fitting a blackbody give similar bolometric corrections, so that at  $19.2$ days after explosion, the bolometric luminosity peaks at 
$ L = 10^{42.2\pm0.1} $ erg $\rm{s}^{-1}$
or $L = 4.12 \times 10^8 L_\odot$. 

\begin{deluxetable}{lll}
%\tablewidth{600pt}
\tablecaption{Some values of variables defined by 
\cite{soderberg} for SN2008D. \label{table4}}
\tabletypesize{\small}

\tablehead{ 
\colhead{Variable} & \colhead{Symbol}  &\colhead{Value} 
}
\startdata
initial radius & $R_*$ & $7 \times 10^{10}$ cm  \\
velocity (phot) & $v_{phot}$ & $11,500$ cm\, s$^{-1}$ \\
velocity (scale) & $v_{sc}$ & $4,743$ cm\, s$^{-1}$ \\
KE & $E_K$ & 2--4 bethe \\
ejected mass  & $M_{ej}$ & 3---5 $M_\odot$ \\
initial $^{56}$Ni   & $M_{Ni}$ & 0.05---0.1 $M_\odot$ \\
\enddata
%\tablenotetext{a}{After shock emergence.}
\end{deluxetable}
\placetable{4}

Table~\ref{table4}
%Table 4 
summarizes parameters derived from \cite{soderberg}.
\cite{modjaz} have further discussed the derived parameters, and most notably found significantly different results for initial radii with shock breakout theories of \cite{W07} or \cite{CF08}.

\begin{figure}[h]
\figurenum{7}
\includegraphics[angle=-90,scale=0.25]{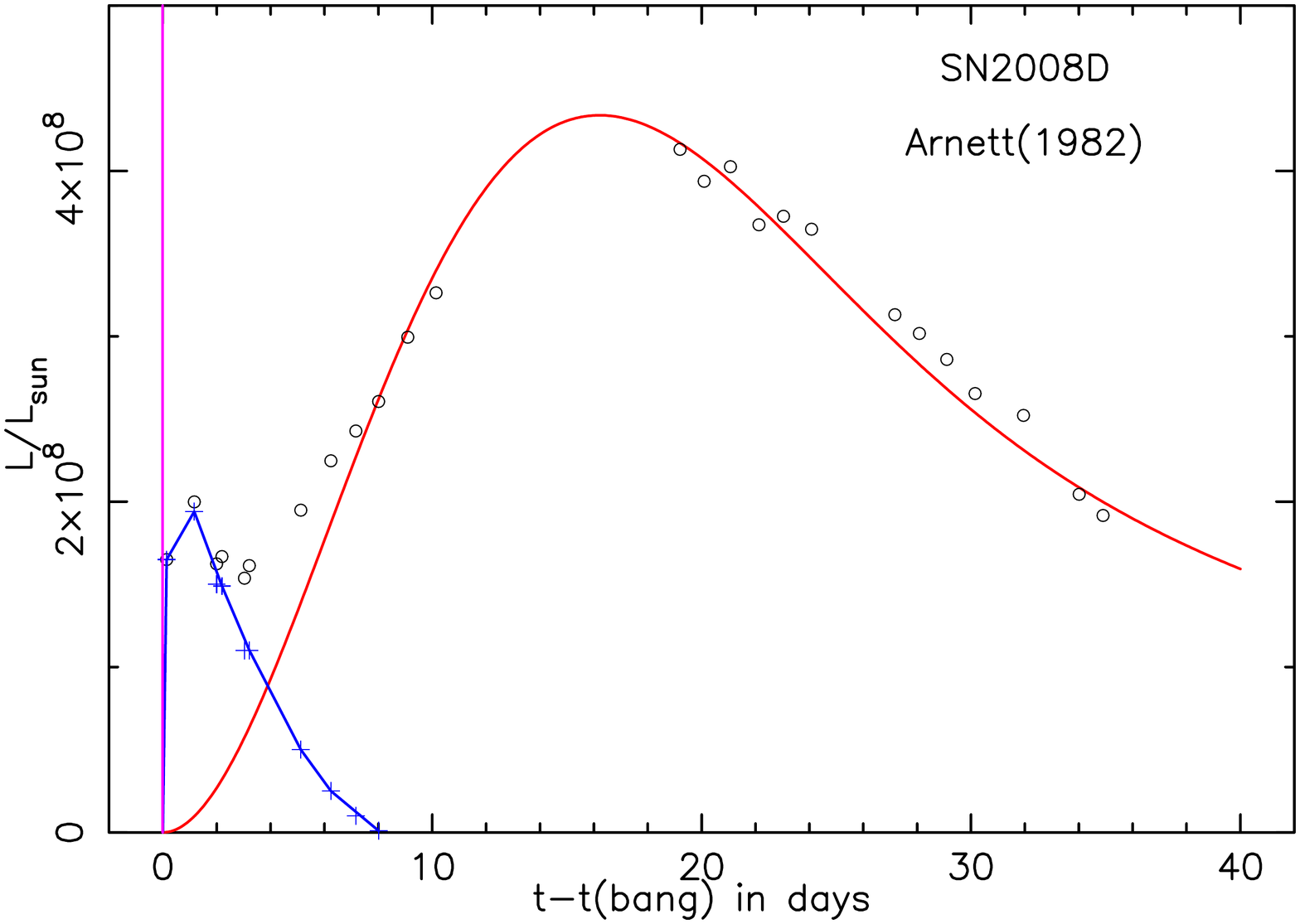}
\caption{Bolometric light curves for SN2008D (circles) compared to analytic models {\em fitted to the luminosity peak at 20 days} (red line). V-magnitudes from this Ib supernova \citep{bianco} are bolometrically corrected using  Fig.\,9 of \cite{modjaz}, and are shown as circles. 
The X-ray spike found by \cite{soderberg} is shown (purple) at the origin in time and allows an accurate determination of explosion time. 
The difference between this purely radioactive part (red line) and the observations (circles) allow us to estimate the excess due to additional processes (blue line and pluses). 
}
\label{SN2008D}
\end{figure}
\placefigure{7}

In order to compare bolometric light curve shapes to observational data, 
we correct the V-band magnitudes \citep{bianco} using graphical extraction of the bolometric corrections from Fig.\,9 in \cite{modjaz}. From 1 to 30 days after outburst, the variation is not large.

In Fig.~\ref{SN2008D} an estimate of the bolometric light curve for SN 2008D is presented. There are three different pieces: (a) an X-ray burst (purple) at explosion time, (b) newly-found, underlying intermediate peak at about 2 days (blue), and (c) a large peak at about 16-20 days (red). We proceed to analyze each piece.

\subsubsection{Main diffusive stage}
The major diffusive part (c) is represented by
an analytic light curve (red) with the parameters given in Table~\ref{table4}, and with  $\kappa_t=0.09$ as used previously for SN2011fe and KSN2008b. For simplicity no attempt was made to correct for effects of varying leakage (opacity).
In this approximation the evolution begins after the shock reaches the surface; there is no shock break-out included. 

\subsubsection{Break-out stage}
The break-out (a) gives a spike in X-rays \citep{soderberg}, followed by cooling of this near-surface region by radiative diffusion and by expansion, quickly merging to the lowest-order diffusion solution \cite{a80,a82a,pe00a}. The break-out spike in luminosity (purple) is much brighter than the  V-band luminosity, because of large bolometric error (the spike radiation begins as X-ray and cools to UV, and so is not UVOIR).
% The shape of the luminosity curve, with an initial spike followed by a broader peak, is a generic feature of supernova models \citep{falk-break,fa73}, and is affected by the initial stellar radius \citep{colgate,a80}. 

\subsubsection{Intermediate peak}
If we subtract the analytic model from the bolometric luminosity, there is an excess which coresponds to a peak  located at $1 < t \leq 8 $ days.
This is piece (b) referred to above and seen in Fig.~\ref{SN2008D} as blue. %, which in fact is indirectly due to shock break-out and cooling.
After this time the total luminosity is well represented by the analytic model (red), which is due only to radioactive decay. 
This intermediate peak (in blue) at $1< t < 8$ days, following the initial X-ray spike, does not appear in SNe Ia. 
This use of the analytic model allows us to probe the underlying physics in more detail.
 
What might cause this intermediate peak? The possibilities are intriguing but not unique.
For example:  

(1) Shock  heating of a clump of matter which was ejected to a radius  $r>10^{14}$\,cm prior to core collapse, perhaps due to pulsations and eruptive mass loss \citep{am16key}, could be responsible. 

(2) A $^{56}$Ni plume of modest mass due to Rayleigh-Taylor instabilities (for example, a mass of $0.028 M_\odot$ in the plume,  with $0.012 M_\odot$ of $^{56}$Ni and $0.3$ bethe of kinetic energy can reproduce the excess luminosity in Fig.~\ref{SN2008D}, with little effect on the main peak). 

(3) Many other possibilities which may be imagined due to activity in the newly-collapsed core (see,  e.g.,  \S\ref{section-derive}).

(4) The breakout might be more complex than now imagined, and relax in a broader peak, not a $\delta$-function.

The first two are likely to give variations from event to event, as might the last two. New instruments such as LSST should clarify the extent of such fluctuations in the light curve of SN Ibc at early times ($t<8$ days). In this paper we have attempted precise comparisons with a few selected data sets of high quality; this has the disadvantage that it may not span the space of natural variation, which may be large for SNe Ibc \citep{jcw-snic}. 

Table~\ref{table4} gives a moderately different set of parameters from Table~\ref{table5} because: (1) the bang time was measured,  (2) the shape of the light curve was fit, including the excess (intermediate peak), and (3) the leakage opacity was assumed the same as in SN2011fe (\S\ref{feLightcurves}). For consistency a small correction for the shift in bang time due to variable leakage (opacity) should be made; 
for simplicity it was not. 
 
The net result of fitting the inferred diffusion peak (blue) separately seems to be a smaller estimate of ejected mass and explosion energy for this SN Ib (compare Table~\ref{table5} to Table~\ref{table4}), but these differences are not drastic.  If the total mass at core collapse is $\sim 3.5 M_\odot$, as we infer from Fig.~\ref{SN2008D}, SN2008D could be the stripped core of a star of initially $12$---$15\, M_\odot$, for example.

\begin{deluxetable}{lll}
%\tablewidth{200pt}
\tablecaption{Parameters for model (red curve) of SN2008D (Fig.~\ref{SN2008D})\label{table5}}
\tabletypesize{\small}

\tablehead{ 
\colhead{Variable} & \colhead{Symbol}  &\colhead{Value} 
}
\startdata
Mass ejected$^a$ & $M_{ej}$ & $2.0\, M_\odot$ \\
Initial $^{56}$Ni & $M_{Ni}(0)$ & $0.082\, M_\odot$ \\
Initial radius & $R(0)$ & $1.0 \times 10^{10} $ cm \\
Velocity scale & $v_{sc}$ & $1.38 \times 10^9$ cm s$^{-1}$ \\
Opacity & $\kappa_t$ & $0.09$ cm$^2$ g$^{-1}$ \\
Explosion energy & $E_{sn}$ & $ 2.3 \times 10^{51}$ ergs \\
%{\color{red}add}\\
\enddata

\tablenotetext{a}{Total stellar mass at core collapse is $\sim 3.5M_\odot$, that is,
 $M_{ej}$ plus the mass of the collapsed core.}

\end{deluxetable}

\section{Conclusion}\label{Conclusion}
 
We have applied the \cite{a82a} model to a wide variety of high-quality data on type Iabc supernovae.
In summary:
\begin{enumerate}
\item There is excellent agreement between the best SNIa bolometric data and analytic models (which are bolometric).
\item There is an important problem of uniqueness; multiple combinations of  parameters may all explain a given light curve. Light curves should be used, but with care, as probes of the supernova engine for core collapse supernovae (SNIbc). A light curve fit might have little to do with details of core collapse.
\item The best very early SNIa data show a deviation from the analytic models, which is easily
removed by the relaxation of the assumption of strictly constant leakage time (opacity) at these times. This also slightly  affects the inferred time from explosion to peak light.
\item Kepler data for KSN2011b may already indicate the presence of turbulent mixing during the explosion. It provides observational support for a slowly  varying leakage time (mean opacity).
\item Arnett's model may be generalized to SNIbc; for SN 2008D, inclusion of radioactive plumes or  pre-collapse eruptions are among the possible causes of excess radiation.
\item Arnett's rule deals with the radioactive heating of the plasma, and so includes effects of gamma-ray escape.
\item The SNe Ib, 2008D, is consistent with the explosion of a stripped-envelope star of initially 12--15$ M_\odot$, possibly with a small radioactive plume.
\end{enumerate}

\acknowledgements
We wish to thank an exceptionally cogent and critical referee for encouragement, and
the Kepler team for providing a machine-readable copy of their data for KSN2011b,c and KSN2012a, 
Federica Bianco for machine-readable data for SN2008D, 
and Gautham Narayan and Charles Kilpatrick for interesting discussions of their work.
We thank the Theoretical Astrophysics Program (TAP) at the University of Arizona, and Steward Observatory for support.

\appendix

\section{Nuclear Data}\label{app_B}

The data for the nuclear decay of $^{56}$Ni and $^{56}$Co in the analytic light curves is documented in Table~\ref{tableA1}.

\begin{deluxetable*}{llcl}
%\tablewidth{600pt}
\tablecaption{Nuclear physics data$^a$\label{tableA1}}
\tabletypesize{\small}

\tablehead{ 
\colhead{Variable} & \colhead{Symbol} & \colhead{Nucleus} &\colhead{Value} 
}
\startdata
Half-life &  $\tau_{1 \over 2}$ &$^{56}$Ni  &   $6.075 $ days \\
 & & $^{56}$Co  &   $77.236 $ days \\
%  & & $^{52}$Fe & $ 8.275 $ hours \\
%  & & $^{52}$Mn & $ 5.591 $ days \\
% &  & $^{48}$Cr & $21.56$ hours \\
% &  & $^{48}$V & $15.9735$ days \\
Total gamma energy  &E${_\gamma}$ & $^{56}$Ni   & 1.750 MeV \\
 & &  $^{56}$Co    & 3.610 MeV\\
%  & & $^{48}$Cr  & 0.634 MeV ??\\
 Positron KE & E$_{\beta^+}$ &  $^{56}$Co    & 0.120 MeV\\
Energy release rates &    $\epsilon_{Ni}$ & $^{56}$Ni & $3.9805 \times 10^{10}$ erg/g/s \\
& $\epsilon_{Co}(\gamma)$ & $^{56}$Co & $ 6.4552 \times 10^9$ erg/g/s \\
& $\epsilon_{Co}(\beta^+)^b$ & $^{56}$Co &  $2.1458 \times 10^8$ erg/g/s \\
\enddata

\tablenotetext{a}{Slightly updated and extended version of \cite{nadyozhin}, using the NNDC Chart of Nuclides \url{www.nndc.bnl.gov/chart}.}

\tablenotetext{b}{Mean kinetic energy of positron emission; positrons are assumed to slow before annihilation, and these gamma rays are assumed to escape.}

\end{deluxetable*}

%\placetable{1}
             
\section{Light curve equations}\label{appendixb}
A more versatile method for solution of the \cite{a82a} model is to  use four coupled ordinary differential equations, which we document below.
\subsection{Equations for Arnett (1982) model}
\begin{enumerate}
\item The first law of thermodynamics,
\begin{equation}
d E/dt + P dV/dt = \epsilon_{deposit} - \partial L/\partial m \nonumber
\end{equation}
becomes, upon integrating over the co-moving spatial variables \citep{a82a} and assuming dominant radiation pressure,
\begin{eqnarray}
    d \phi(t) / dt =  R(t)/R(0)  \nonumber\\
     \Big (  M_{Ni} \Big [ ( \epsilon_{Ni} X_{Ni}+ \epsilon_{Co} X_{Co} ) D \nonumber\\
      +\epsilon_{Co}^+ X_{Co} \Big ]    - \phi(t) /\tau_{dif} \Big ), \nonumber
\end{eqnarray}
where $D$ is the deposition function of \cite{cpk80} for $\gamma$-rays, and  $\tau_{dif} = \kappa_t M_{ej} /(13.7 c R(0))$, and
$\phi(t) = a [(T(t)R(t))/T(0)R(0)]^4$.
This is mathematically equivalent to Eq.~1 of \cite{kkd13}.

\item Constant homologous expansion (Hubble flow):
\begin{equation}  
d R / dt = v_{sc} ,\nonumber
\end{equation}
where
 $v_{sc} = \Big ( 10 E_{KE}/ 3 M_{ej} \Big )^{1 \over 2},$
\item $^{56}$N decay:
\begin{equation}  
      d X_{Ni} /dt = - X_{Ni}/ \tau_{Ni}, \nonumber
\end{equation}
\item $^{56}$Co decay:
\begin{equation}
d X_{Co} /dt = -d X_{Ni}/dt - X_{Co}/\tau_{Co},  \nonumber
\end{equation}

\end{enumerate}
See also Table~\ref{tableA1}.

These equations are integrated by Runge-Kutta methods to a global accuracy of better than one part in $10^8$.
For the same physical assumptions concerning escape of $\gamma$-rays, i.e., the function $D$, these equations and  the integral approach of \cite{kkd13} should give consistent results (fitting the light curve with this approach may give more information, as seen in \S\ref{sn2008D}).

\subsection{Changes and extensions}
To include another radioactivity, add a new decay equation (like 3 annd 4), and add the new plasma heating to 1.

To add a new source of energy (magnetar, accretion disk, etc.), define a plasma heating rate as a function  of time and add to 1.

To do energy tests or a \cite{kkd13} analysis, add new equations and variables (integrands) to integrate the energy types over time.

Typos do happen, but try to validate and verify before publishing (e.g., reproduce Figure~\ref{fig_sn2011fe}  for the same parameters).

\end{document}